\documentclass[prb,reprint,aps,showpacs,ams,twocolumn]{revtex4} 
\usepackage{graphicx} 
\usepackage{dcolumn} 
\usepackage{latexsym}
\usepackage[dvips]{color} 

\begin{document} 
 
\title{ 
Multiple-gap structure 
in electric-field-induced surface superconductivity
} 
\author{Yousuke Mizohata} 
\affiliation{Department of Physics, Okayama University, 
Okayama 700-8530, Japan}  
\author{Masanori  Ichioka} 
\affiliation{Department of Physics, Okayama University, 
Okayama 700-8530, Japan} 
\author{Kazushige  Machida} 
\affiliation{Department of Physics, Okayama University, 
Okayama 700-8530, Japan} 
\date{\today}

\begin{abstract}
Local superconducting gap structure is studied 
as a function of nanoscale depth 
in electric-field-induced surface superconductivity  
such as in ${\rm SrTiO_3}$. 
We examine solutions of Bogoliubov-de Gennes equation 
in two limiting confinement potential cases of electric field 
with and without screening effects. 
As unique properties different from bulk superconductivity, 
there appear in-gap states 
even for isotropic $s$-wave pairing, 
due to multiple gap structure of sub-band dependent 
surface superconductivity.  
These determine the depth-dependence of local superconductivity.
\end{abstract}
 

\maketitle 
\section{Introduction} 

Electric-field-induced carrier-doping technique,  
using field-effect-transistor (FET) structure or 
electric-double-layer-transistor (EDLT) 
structure,~\cite{Ueno,Ye,UenoKTO,Taniguchi,YeMoS2} 
attracts much attention 
as a new method to carrier doping, other than the methods of 
chemical doping. 
A merit of electric-field-induced doping is that 
we can control the doping carrier density by a gate voltage in a same sample.  
This will be a powerful platform 
in future studies of condensed matter physics. 
When this is used at the surface of insulators, 
carriers are induced near the surface, and trapped  
in the confinement potential of the electric field. 
Using enough strong field by EDLT, 
we can realize superconductivity of the surface metallic states 
at low temperature $T$, such as in 
${\rm SrTiO_3}$,~\cite{Ueno} 
${\rm ZrNCl}$,~\cite{Ye} 
${\rm KTaO_3}$,~\cite{UenoKTO} 
${\rm MoS_2}$.~\cite{Taniguchi,YeMoS2} 
The gate voltage control of surface superconductivity in ${\rm SrTiO_3}$ 
was also realized at the interface of 
${\rm LaTiO_3 / SrTiO_3}$
and ${\rm LaAlO_3 / SrTiO_3}$.~\cite{Biscaras,Bert} 

Compared to these developments of experimental research, 
theoretical understandings are not enough 
for properties of the electric-field-induced surface superconductivity.
We have to discuss whether the surface superconductivity has 
the same properties to those of bulk superconductivity, 
or whether it has different unique properties. 
In future experiments for unconventional superconductivity 
produced on a surface, 
we have to distinguish unique properties of surface superconductivity 
and exotic properties of unconventional superconductivity. 

As different properties from bulk metallic states, 
sub-bands are formed in the surface metallic states 
due to the confinement potential by strong electric fields.~\cite{Ueno} 
Since multiple sub-bands are occupied by surface carriers,  
this system is not ideal two-dimensional states.  
We also note that the local carrier density $n(z)$ has spatial variation 
as a function of depth $z$ from the surface 
in the surface metallic states, 
while $n(z)$ is constant in bulk metallic states. 
The quantitative estimate for the $z$ dependence is 
one of problems for the electric-field-induced metallic state. 
Therefore, also in the theoretical studies of surface superconductivity, 
we need to know detailed spatial structure of the superconducting gap 
in the nanoscale and its sub-band dependence. 
These studies enable us to find 
differences from bulk superconductivity. 

In this letter, we study unique properties; 
local electronic states and sub-band dependence   
in electric-field-induced superconductors.  
We will discuss multiple-gap structure of 
the sub-band dependent surface superconductivity.  
Since we determine the spatial structure in the order of 
Thomas-Fermi length near the surface, 
we solve the Bogoliubov-de Gennes (BdG) equation~\cite{BdG}  
under the electric-field $F(z)$. 
We discuss the depth $z$ dependence perpendicular to the surface at $z=0$. 
As for confinement potential $V(z)$ by $F(z)$, we compare two cases;  
triangular potential and self-consistent potential.~\cite{Stern,Ando} 
The latter is the case 
when induced carriers completely screen the applied electric field. 
The former is the opposite limit where the screening is negligible. 

This paper is organized as follows. 
After we explain our theoretical formulation 
of BdG equation under electric fields in Sec. \ref{sec:formulation}, 
we study the depth-dependence of local superconducting gap structure 
in Sec. \ref{sec:LDOS}, and 
the gap structure of the sub-band modes in the spectral weight  
in Sec. \ref{sec:SW}. 
In order to discuss the relation of sub-band dependent gap structure 
and the depth-dependence of superconducting states, 
we perform  the analyses of sub-band decomposition 
for the surface superconductivity in Sec. \ref{sec:subband}. 
The last section devotes to discussion and summary, 
including the topics of 
the Bardeen-Cooper-Schrieffer(BCS) - Bose Einstein condensation(BEC) 
crossover phenomena in the surface superconductivity.

\section{Bogoliubov-de Gennes theory in confinement by  electric field} 
\label{sec:formulation}

Throughout this letter, energy, length, local carrier densities are, 
respectively, presented in unit of ${\rm eV}$, ${\rm nm}$, 
and ${\rm nm}^{-3}$. 
We typically consider the case of  
sheet carrier density 
$n_{\rm 2D}=6.5 \times 10^{13}$[${\rm cm^{-2}}$],
and electric field at the surface is given by 
$F_0 \equiv F(z=0)=1.4 \times 10^{-3}$[${\rm V/nm}$]. 
The triangular potential with this $F_0$ corresponds to 
one of the case calculated in Ref. \onlinecite{Ueno} for ${\rm SrTiO_3}$, 
while single band case of effective mass $m^\ast=4.8m_0$ 
is considered here. $m_0$ is free electron's mass. 

In the normal state,~\cite{Ueno,Stern} 
the eigen-energy $E_\epsilon$ and wave function 
$u_{\epsilon}({\bf r}) 
={\rm e}^{{\rm i}(k_x x + k_y y)} u_{\epsilon}(z) /\sqrt{S} $ 
are determined by the Schr\"{o}dinger equation 
\begin{eqnarray}
K u_{\epsilon}(z)  =E_{\epsilon} u_{\epsilon}(z) ,
\label{Sch}
\end{eqnarray}  
with 
\begin{eqnarray} 
K=-\frac{\hbar^2}{2m*}\frac{{\rm d}^2}{{\rm d}z^2}+E_\parallel +V(z)-\mu,
\label{Sch2}
\end{eqnarray}  
$E_\parallel=\hbar^2 k^2_\parallel / 2m^\ast $, 
$k^2_\parallel=k_x^2+k_y^2$, and $S$ is unit area of surface. 
We assume that the wave functions vanish at $z=0$ as the boundary condition.
In the parallel direction to the surface, the eigen states are given by 
wave numbers $k_x$ and $k_y$ of plane waves.  
Thus, the eigen-states of Eq. (\ref{Sch}) are 
labeled by $\epsilon \equiv (k_x,k_y,\epsilon_z)$. 
$\epsilon_z$ indicates label for sub-bands coming from quantization by 
confinement in the $z$-direction. 
The local carrier density is calculated as 
$n(z)=2n_{\uparrow}(z)$ with 
\begin{eqnarray} && 
n_{\uparrow}(z)=\sum_{\epsilon} |u_{\epsilon}(z)|^2 f(E_{\epsilon}), 
\label{nz}
\end{eqnarray} 
where $f(E)$ is the Fermi distribution function. 
To fix $n_{\rm 2D}$,  
we tune chemical potential $\mu$.
In the triangular potential case, 
the confinement potential is given by 
$V(z)=|e|F_0 z$. 
For the selfconsistent potential, 
\begin{eqnarray} 
F(z)=F_0 \left( 
1 - \int_0^z n(z'){\rm d}z' / n_{\rm 2D} \right), 
\label{Fz}
\end{eqnarray} 
by Gauss's law, 
considering the screening by $n(z)$, 
and 
\begin{eqnarray}
V(z)=|e|\int_0^z F(z'){\rm d}z'. 
\label{Vz}
\end{eqnarray}
As $F(z \rightarrow \infty )=0$, 
$ n_{\rm 2D} =\int_0^\infty n(z'){\rm d}z'$.  
Iterating calculations of 
Eqs.(\ref{Sch})-(\ref{Sch2}) 
and Eqs.(\ref{nz})-(\ref{Vz}) 
in the region $0 \le z \le L$, 
we determine $V(z)$ 
in the case of selfconsistent potential. 
Typically we use $L=80$[nm].  

In the superconducting state, 
the wave function 
\begin{eqnarray} 
\left(\begin{array}{c}
u_{\epsilon}({\bf r}) \\ 
v_{\epsilon}({\bf r}) \\ 
\end{array}\right) 
=
\frac{1}{\sqrt{S}}{\rm e}^{{\rm i}(k_x x + k_y y) }
\left(\begin{array}{c}
u_{\epsilon}(z) \\ 
v_{\epsilon}(z) \\ 
\end{array}\right) 
\end{eqnarray} 
is determined by solving the BdG equation~\cite{BdG,Takigawa,Takahashi}   
\begin{eqnarray} 
\left(\begin{array}{cc}
K & \Delta(z) \\ 
\Delta(z) & -K \\ 
\end{array} \right) 
\left(\begin{array}{c}
u_{\epsilon}(z) \\ 
v_{\epsilon}(z) \\ 
\end{array}\right) 
=E_{\epsilon} 
\left(\begin{array}{c}
u_{\epsilon}(z) \\ 
v_{\epsilon}(z) \\ 
\end{array}\right) . 
\end{eqnarray} 
The pair potential $\Delta(z)$ is selfconsistently calculated by 
\begin{eqnarray}
\Delta(z)
=V_{\rm pair} \sum_{\epsilon}
  u_{\epsilon}(z) v_{\epsilon}(z) f(-E_{\epsilon}) 
\label{eq:Delta}
\end{eqnarray}
with  
the energy cutoff $ E_{\rm cut} $
of the pairing interaction. 
Here, we consider a conventional case 
of isotropic $s$-wave pairing. 
We typically use $V_{\rm pair}=0.04$, $E_{\rm cut}=0.01$,  
and $T \sim 0$. 

\section{Depth-dependence of local superconducting gap structure} 
\label{sec:LDOS}

\begin{figure}[tb] 
\begin{center}
\includegraphics[width=9.0cm]{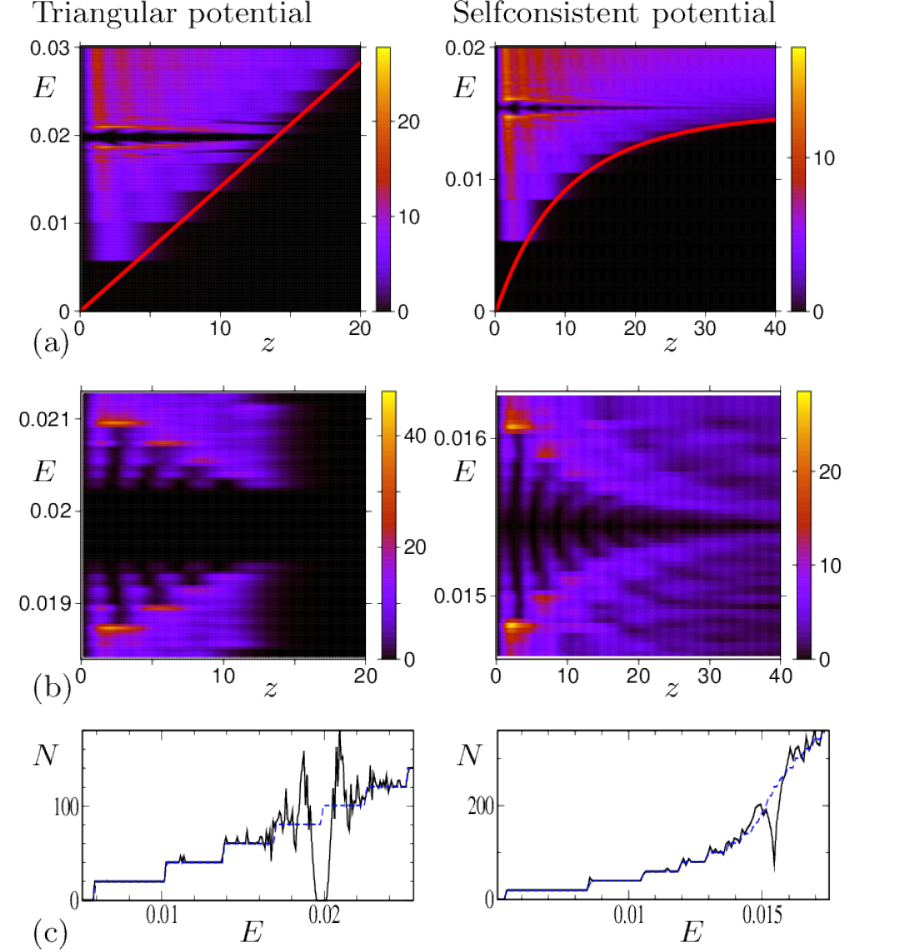} 

\vspace{-0.5cm}
\end{center}
 \caption{ 
(Color online)  
(a) Density plot of LDOS $N(E,z)$ as a function of $z$ and $E+\mu$. 
Solid line presents the confinement potential $V(z)$. 
(b) LDOS $N(E,z)$ in (a) is focused near the superconducting gap.
(c) DOS $N(E)$ as a function of $E+\mu$.  
Dashed lines present $N(E)$ for normal state. 
Left (Right) panels are for triangular potential (selfconsistent potential).  
} 
\label{fig1} 
\end{figure} 

First, we study the 
local density of states (LDOS) 
$N(E,z)=2N_{\uparrow}(E,z)$ with~\cite{Takigawa,Takahashi}    
\begin{eqnarray} && 
N_{\uparrow}(E,z)
=\sum_{\epsilon=(k_x,k_y,\epsilon_z)} |u_{\epsilon}(z)|^2 \delta(E-E_{\epsilon}) . 
\end{eqnarray} 
The left panel of Fig. \ref{fig1}(a) presents 
$N(E,z)$ for the triangular potential. 
There, we see steps of LDOS by the sub-band structure 
of quantized bound states, 
as in the normal state.~\cite{Ueno}  
The lowest sub-band appears at $E> E_{l=1,{\rm min}} \sim 0.0059$ 
near the surface. 
The continuum distribution above $E_{l=1,{\rm min}}$ comes 
from finite $E_\parallel \ge 0$. 
Similarly there appears the LDOS of second sub-band 
at $E>E_{l=2,{\rm min}}\sim 0.0103$,  
and the LDOS of third sub-band 
at $E>E_{l=3,{\rm min}} \sim 0.0138$. 
Their contributions are overlapped each other at higher energies. 
When the sub-band level $l$ is higher, 
the eigen-energy $E_{l,{\rm min}}$ becomes higher, 
and the distribution spread until deeper $z$ from the surface. 
The superconducting gap appears near $\mu \sim 0.0198$. 

The $N(E,z)$ for the self-consistent potential 
is presented in the right-panel of Fig. \ref{fig1}(a). 
There,  we see 
step-structures 
of sub-bands at low energy region, 
but the step size becomes smaller at higher energy, 
because the slope of 
$V(z)$ decreases to zero 
as a function of $z$ by the screening effect. 
Since the chemical potential is located at 
$\mu \sim V(z \rightarrow \infty) \sim 0.0154$, 
occupied states with $E < \mu$ are bound states, 
and empty states with $E>\mu$ are scattering states 
which are free from the confinement potential. 
The superconducting gap opens between the bound states 
and the scattering states. 

The superconducting gap structures are focused in Fig. \ref{fig1}(b). 
Even in the isotropic $s$-wave pairing, 
we see in-gap states which have oscillations 
as a function of $z$ and steps of gap-edges as a function of $E$, 
as characteristic features of electric-field-induced surface 
superconductivity. 
High intensity peaks of $N(E,z)$ correspond to the maximum gap-edge, 
whose gap amplitude decreases discontinuously with increasing $z$. 
In the selfconsistent potential (right-panel),  
it reduces to zero at large $z$. 

Figure \ref{fig1}(c) shows density of states (DOS) $N(E)$ 
after $z$-integration of $N(E,z)$.  
Because of the in-gap states, 
gap structure in $N(E)$ is different 
from that of bulk isotropic $s$-wave superconductors.  
In the triangular potential (left panel), 
the gap-edge has width from minimum gap to gap-edge peak of maximum gap, 
as in anisotropic $s$-wave superconductors.  
In the selfconsistent potential (right panel), 
full-gap structure does not exist, 
since low energy states exist until near $\mu$.  
The gap shape is similar to that 
of anisotropic superconductors with gap nodes. 

\begin{figure}[tb] 
\begin{center}
\includegraphics[width=9.0cm]{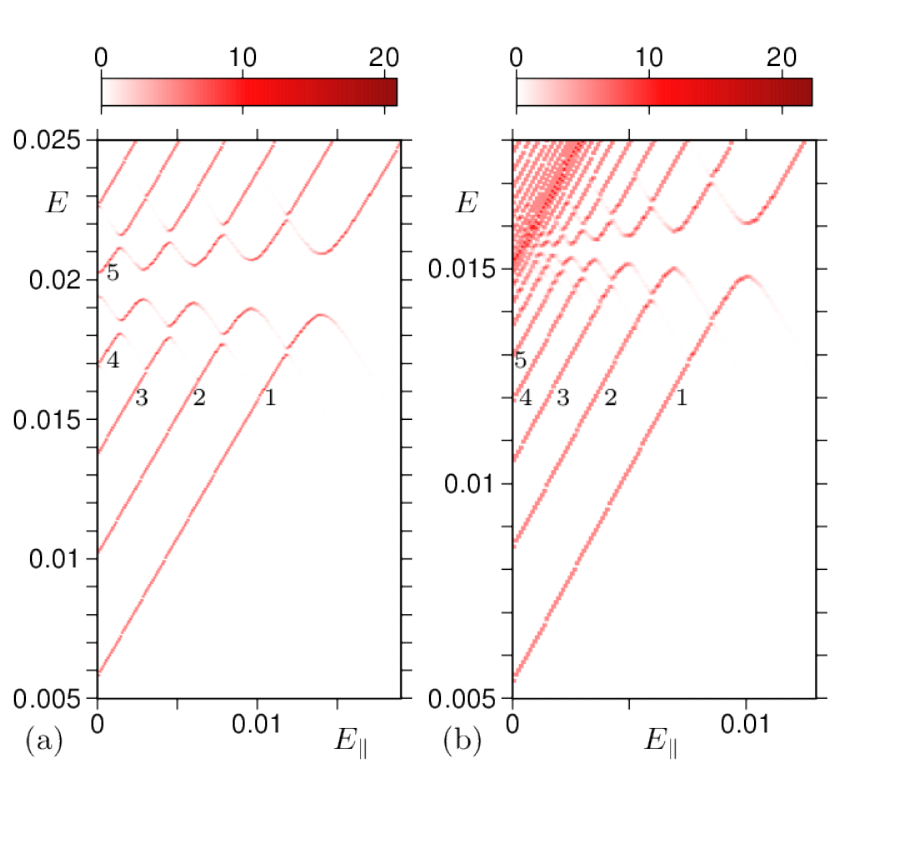} 

\vspace{-1.5cm} 
\end{center}
 \caption{ 
(Color online)  
Density plot of spectral weight $N(E,k_\parallel)$ 
as a function of $E_\parallel=\hbar^2 k_\parallel^2 /2m^\ast$ and 
$E+\mu$ for (a) triangular potential and (b) selfconsistent potential.  
Numbers in the figure indicate sub-band level $l$.
} 
\label{fig2} 
\end{figure} 
\section{Gap structure of sub-band mode in  spectral weight} 
\label{sec:SW}

We discuss that these superconducting gap structures 
come from the sub-band dependence of superconductivity. 
For the sub-band decomposition, 
we calculate the spectral weight 
$N(E,k_\parallel,z)=2N_{\uparrow}(E,k_\parallel,z)$ given by  
\begin{eqnarray} && 
N_{\uparrow}(E,k_\parallel,z)
=\sum_{\epsilon_z} |u_{\epsilon}(z)|^2 \delta(E-E_{\epsilon}) 
\label{eq:sw}
\end{eqnarray} 
from 
$N_{\uparrow}(E,k_\parallel,z)=-{\pi}^{-1}{\rm Im}G_\uparrow(E,k_\parallel,z)$  
with Green's function 
\begin{eqnarray} &&
G_\uparrow(E,k_\parallel,z)
=\int {\rm e}^{-{\rm i}(k_x \tilde{x}+k_y \tilde{y})} 
 G_\uparrow(E,{\bf r},{\bf r}'){\rm d}\tilde{x}{\rm d}\tilde{y}|_{z=z'},  
\quad
\\ && 
G_\uparrow(E,{\bf r},{\bf r}')
=\sum_\epsilon 
\frac{u^\ast_\epsilon({\bf r}) u_\epsilon({\bf r}')}{E+{\rm i}0-E_\epsilon}, 
\end{eqnarray}
and $(\tilde{x},\tilde{y}) \equiv ({x-x'},{y-y'})$.~\cite{Ichioka} 
The $z$-integration of $N(E,k_\parallel,z)$ is given by 
$N(E,k_\parallel) =\int_0^L N(E,k_\parallel,z){\rm d}z$.  
In Fig. \ref{fig2} 
we show 
$N(E,k_\parallel)$, 
which appear at the eigen energies $E_\epsilon$.  
There we see multiple parallel lines of the dispersion relation 
as a function of $E_\parallel$, 
corresponding sub-bands of surface bound states. 
From the bottom, the lines are assigned to sub-band level $l=1$, 2, $\cdots$, 
as indicated in Fig. \ref{fig2}.   
In the case of self-consistent potential, 
the energy distance of the dispersion relation between sub-bands decreases 
for higher sub-bands, and the spectral weight becomes continuous 
near $E_\parallel \sim 0$ at $E>\mu$ in the scattering state.   

In the superconducting state, gaps open at 
crossing points of the particle mode and the inverted hole mode at $E=\mu$,   
forming Bogoliubov's dispersion relations 
of superconductivity for each level of sub-band. 
The superconducting gap is larger for lower sub-bands, 
indicating multiple-gap structure of surface superconductivity. 
In the case of triangular potential, 
the occupied lower sub-bands have different but finite gaps. 
In the case of self-consistent potential, 
sub-bands are occupied until quite higher levels, 
where superconducting gap reduces to zero. 
Therefore, the in-gap states appear until near $E = \mu$. 
As for the $z$-dependence of spectral weight $N(E,k_\parallel,z)$,  
the contribution of lower sub-band is dominant near the surface. 
The contributions of higher sub-bands become dominant at deeper $z$.

\begin{figure}[tb] 
\begin{center}
\includegraphics[width=8.5cm]{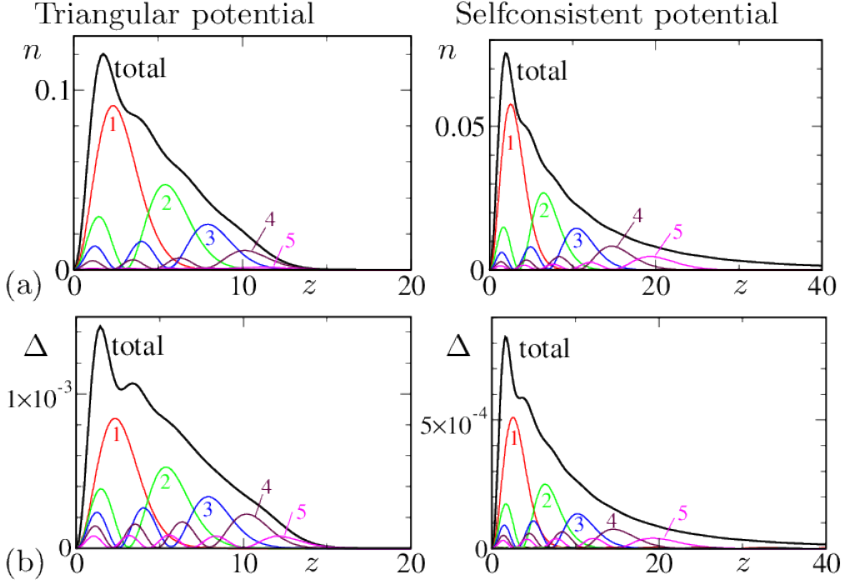} 

\vspace{-0.5cm}
\end{center}
 \caption{ 
(Color online)  
(a) Local carrier density $n(z)$ and 
(b) superconducting pair potential $\Delta(z)$ 
as a function of $z$. 
The sub-band decompositions are 
also presented for sub-band levels $l=1,\cdots,5$. 
Left (Right) panels are for triangular potential (selfconsistent potential).  
} 
\label{fig3} 
\end{figure} 

\section{Sub-band decomposition of local superconducting states} 
\label{sec:subband}

In Fig. \ref{fig3}(a), 
we present the local carrier density $n(z)$  
and the sub-band decomposition. 
The eigen-states on the dispersion relations in Fig. \ref{fig2} 
are classified to each sub-band level $l$. 
In $l$-th sub-band,  
the wave function of the form of Airy functions 
has $l-1$ nodes along $z$ direction.~\cite{Ueno} 
The higher sub-band contributions can penetrate into deeper $z$. 
Since the LDOS is integrated over 
$E_{l,{\rm min}}<E<\mu$ to obtain $n(z)$,  
lower sub-band contributions to $n(z)$ becomes larger, 
because of smaller $E_{l,{\rm min}}$.    
The pair potential $\Delta(z)$ and 
the sub-band decomposition in 
Fig. \ref{fig3}(b) have similar spatial structure 
to those of $n(z)$. 
It is noted that sub-band dependent pair potential becomes smaller 
for higher sub-bands. 
In the selfconsistent potential (right panels), 
while lower sub-band contributions are dominant, 
$n(z)$ and $\Delta(z)$ includes contributions from 
further higher sub-band levels $l>5$. 
Therefore, tails of $n(z)$ and $\Delta(z)$ survive until deeper $z$. 

\begin{figure}[tb] 
\begin{center}
\includegraphics[width=8.5cm]{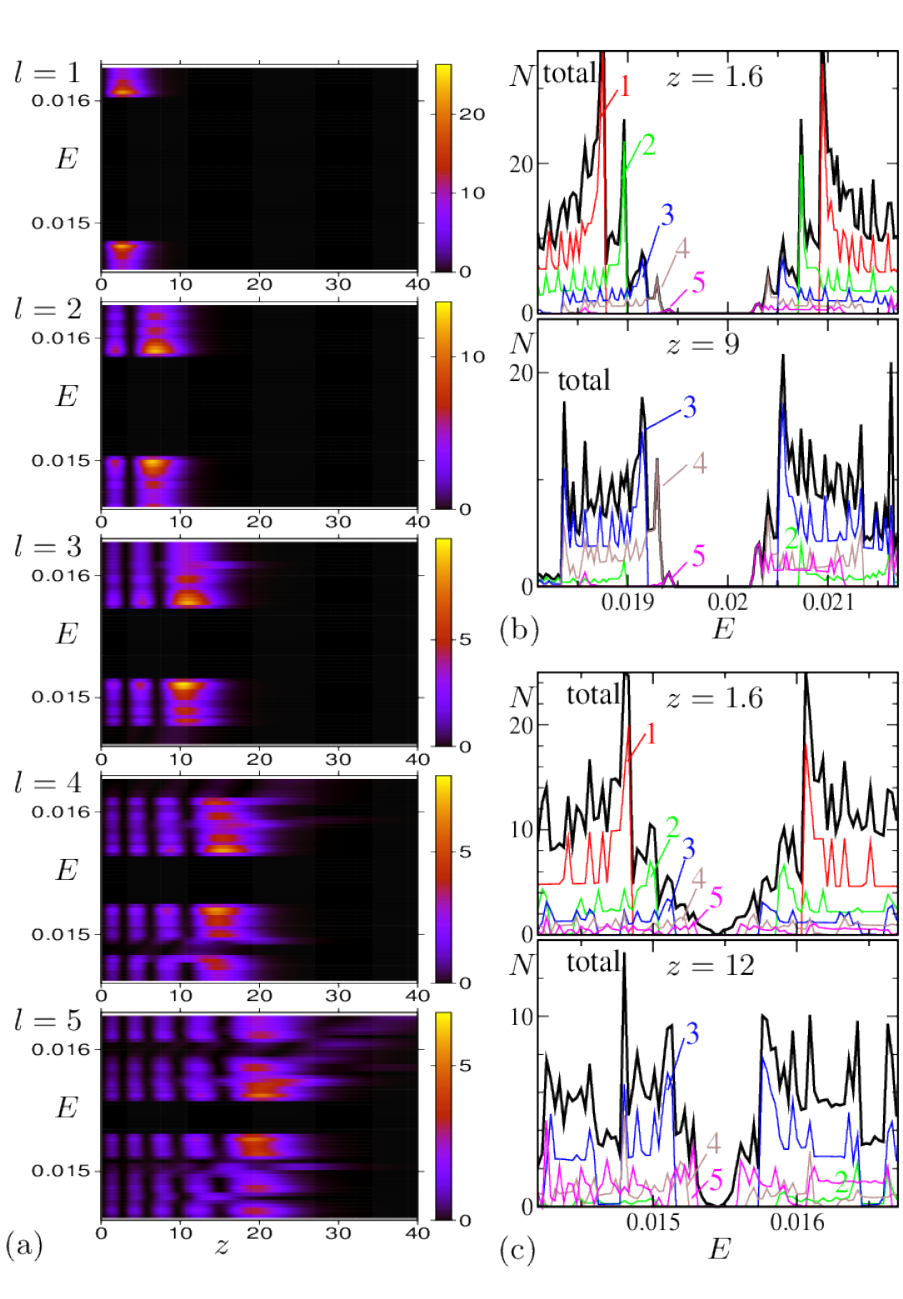} 

\vspace{-1.0cm}
\end{center}
 \caption{ 
(Color online)  
(a) 
Density plot of sub-band decomposition of LDOS  
as a function of $z$ and $E+\mu$ 
for sub-band levels $l$=1, 2, $\cdots$, 5   
in the case of self-consistent potential. 
(b)  
LDOS $N(E,z)$ and the sub-band decomposition 
at $z=1.6$ and 9 in the case of triangular potential.
(c)  
The same as (b), but 
at $z=1.6$ and 12 in the case of selfconsistent potential.
} 
\label{fig4} 
\end{figure} 

To discuss the origin of superconducting gap structure in Fig. \ref{fig1}, 
sub-band decompositions of the LDOS are presented in Fig. \ref{fig4}. 
From Fig. \ref{fig4}(a), 
we see that lowest sub-band contribution ($l=1$) to $N(E,z)$ 
has large constant superconducting gap, 
but its distribution is restricted in very near the surface. 
The contributions from higher sub-bands $l=2$, 3, $\cdots$ have  
smaller constant gap, 
and the distributions spread until deeper $z$.   
By the combination of these sub-band contributions, 
the in-gap states and the $z$-dependence of superconducting gap structure 
in Fig. \ref{fig1} are created. 
These sub-band contributions are clear also in 
the LDOS spectra in Figs. \ref{fig4}(b) and \ref{fig4}(c).   
There, we see multiple peaks of gap edge from sub-band contributions. 
Near the surface ($z=1.6$), all sub-band contributions appear, 
and lower sub-band contributions are dominant. 
Therefore the main peak corresponds to gap edge of largest gap 
by the lowest sub-band.  
In lower panels of Figs. \ref{fig4}(b) and \ref{fig4}(c) for deeper $z$, 
since lower sub-band contributions ($l=1$ and 2) vanish, 
the main peak of gap edge appears at lower gap 
energy corresponding to higher sub-band ($l=3$) contributions.  

In addition to the superconducting gap at $E=\mu$, 
there appear extra small gaps at energies outside of superconducting gap, 
as seen in Fig. \ref{fig2}. 
This occurs by the crossing of the hole- 
and particle-modes between different sub-bands.  
We see these extra gaps 
in higher sub-band contributions also in Fig. \ref{fig4}(a). 
Because of the extra gaps, 
the LDOS in Figs. \ref{fig4}(b) and \ref{fig4}(c) 
has many extra peaks outside of the main gap energy. 

We note that low-energy in-gap electronic states 
are not determined locally by $\Delta(z)$ 
in the length order of nano-scale in this system. 
This is contrasted to conventional case 
when $\Delta(z)$ is suppressed in the length scale of 
superconducting coherence 
length.~\cite{deGennesSaintJames,Tanaka,Im,Kashiwaya}   
There low-energy states appear as localized states 
by the suppression of the local gap. 
In the system of the electric field-induced surface superconductivity, 
approaching $z \rightarrow 0$ near the surface, 
$\Delta(z)$ is suppressed toward zero 
in the length order of nano-scale, 
as shown in Fig. \ref{fig3}(a). 
However, from Fig. \ref{fig4}(a), we see that 
localized low energy in-gap states do not appear  
at the surface region ($z<2$) of suppressed $\Delta(z)$. 
The local state of lowest sub-band has uniform gap with 
largest gap amplitude [top panel in Fig. \ref{fig4}(a)]. 
This indicates that the in-gap states reported in this paper  
is not due to the suppression $\Delta(z \rightarrow 0)\rightarrow 0$. 
Rather, the in-gap states comes from the deeper region,  
as tails of wave functions for higher sub-band levels 
in Fig. \ref{fig4}(a). 
This is one of intrinsic natures 
in the electric field-induced superconductivity.

\section{Discussion and summary}

As future experiments to confirm the in-gap states 
due to the characteristic multiple gap structure, 
we expect observations of LDOS 
such as by point contact tunneling spectroscopy,  
which will see the gap structure in upper panels of Figs. 4(b) and 4(c). 
The contributions of the in-gap states will be 
observed in experiments sensitive 
to DOS of the superconducting gap structure, 
such as magnetic resonance, 
optical absorptions, etc. 
Electric-field-induced doping will be important platform 
to study unconventional superconductivity. 
Before that, it is important to clarify 
the difference of properties between 
surface superconductivity and bulk superconductivity 
in conventional superconductors, as suggested in this work. 
As a concept of multiple-gap structure, 
electric-field-induced surface superconductivity can be said 
a new type of multi-band superconductors. 
We will see some similar behaviors to those of  
multi-band superconductors such as in ${\rm MgB_2}$ 
and Fe-based superconductors.  
The number of contributing sub-bands can be controlled by the gate voltage. 

We point out an interesting possibility to realize the 
BCS-BEC crossover phenomena~\cite{Eagles,Leggett,Nozieres} 
by controlling the gate voltage in surface superconductivity. 
In the cold atomic gases, the BCS-BEC crossover is seen 
by tuning the interaction via a Feshbach resonance.~\cite{Zwierlein,Jochim} 
The BCS-BEC crossover in multi-band superconductor was suggested by 
ARPES experiment in ${\rm FeSe_xTe_{1-x}}$.~\cite{Lubashevsky} 
The same situation appears in the surface superconductivity. 
In Fig. \ref{fig2}(a), 
the superconducting gap in 5th sub-band 
opens at the bottom of the band dispersion. 
That is, since the gap amplitude $|\Delta|$ is larger than the Fermi energy 
$E_{\rm F} (\equiv \mu - E_{l=5,{\rm min}})$ from the band bottom, 
the BEC regime $|\Delta|>E_{\rm F}$ is realized. 
The gaps in other 1-4 th bands are in 
the BCS regime $|\Delta|<E_{\rm F}$. 
As mentioned above, the gap magnitude can be tuned 
by the gate voltage. 


In summary, 
local superconducting gap structure 
and the sub-band dependence 
in electric-field-induced surface superconductivity 
were studied by solving microscopic BdG equation. 
There, the in-gap states appears due to the multiple gap structure 
of multiple sub-band superconductivity, even for isotropic $s$-wave pairing. 
We evaluated how these structures depend on the screening condition, 
i.e., triangular potential or selfconsistent potential. 
These different characters from the bulk superconductivity, 
by the sub-band dependent multi-gap nature, 
are important to be considered, 
when we discuss properties of 
electric-field-induced surface superconductivity.

\section*{Acknowledgements}

We thank Profs. Y. Iwasa, K. Ueno, and T. Nojima for 
fruitful discussions, and information of their 
experimental results.
This work was supported by KAKENHI (No. 21340103).

 
\end{document}